# High-Density Multi-Depth Human Recordings Using 45 mm Long Neuropixels Probes

Daril E. Brown II, PhD,[1] Elizaveta Okorokova, PhD,[1] Carrina Iacobacci, BS,[1] Brian Coughlin,[2,3] Orin Bloch, MD,[1] Eric M. Trautmann, PhD,[1] Sydney S. Cash, MD, PhD,[2,3] Angelique C. Paulk, PhD,[2,3] Sergey D. Stavisky, PhD,*[1] and David M. Brandman, MD, PhD,*[1]

**Affiliations:** [1] Department of Neurological Surgery, University of California Davis Health, Davis, CA, USA. [2] Center for Neurotechnology and Neurorecovery, Department of Neurology, Massachusetts General Hospital, Boston, MA, USA. [3] Department of Neurology, Harvard Medical School, Boston, MA, USA.
*co-senior authors

# Abstract

OBJECTIVE

Neuropixels probes, initially developed for use in small animal models, have transformed basic neuroscience by enabling high-density, single-cell resolution recordings across multiple brain regions simultaneously. The recent development of Neuropixels 1.0 NHP Long—a longer probe designed for non-human primates—has expanded this capability, enabling unprecedented simultaneous access to multiple cortical layers and deep brain structures of large-brained animals. This probe features 4,416 recording sites along a 45 mm shank, with 384 channels selectable for simultaneous recording. Here, we report the first use of these probes in humans, aiming to establish safe intraoperative use and assess feasibility for clinical and research applications.

METHODS

Nine patients undergoing neurosurgical procedures—including epilepsy or tumor resection and deep brain stimulation (DBS) implantation—were enrolled. The authors developed sterilizable, custom-designed 3D-printed tools and protocols to facilitate long Neuropixels probe insertion, optimize recordings, and maintain sterility. Strategies were implemented to mitigate potential failure modes, including motion artifacts and electrical interference.

RESULTS

Successful intraoperative recordings were obtained from surface and deep cortical structures without probe breakage or adverse events. Compared with conventional electrodes, the Neuropixels probe enabled dense sampling across multiple parenchymal depths with submillisecond temporal resolution. Recordings were obtained from deep targets including the hippocampus (n = 3) and cingulate cortex (n = 1), as well as from regions that are challenging to access with single-unit precision, such as the superior frontal sulcus (n = 1). Custom tools and refined workflows lowered technical barriers for operative use and improved recording stability. Neural activity was observed across all recordings.

CONCLUSIONS

Neuropixels 1.0–NHP Long probes can be deployed in the human operating room, enabling simultaneous recordings from multiple brain structures at single-neuron resolution. These methods expand opportunities for studying human brain function and pathology in vivo, and may ultimately support the development of more precise neurosurgical interventions.

# Introduction

Recent advances in neural recording technologies have enabled high-density, single-cell resolution electrophysiological recordings of potentially hundreds of neurons from the human brain, offering unprecedented insights into its function[1–4] and pathology[2,5,6]. Among these tools, Neuropixels probes[7]—silicon-based electrode arrays with thousands of recording sites—offer spatial and temporal resolution far beyond that of clinically approved devices and even existing, commonly used microelectrode systems. While these probes have transformed systems neuroscience in animal models[8–12], their use in humans remains nascent.

Neuropixels have enabled simultaneous recordings across multiple brain regions in rodents[9–12], non-human primates[13], and songbirds[14], revealing insights into the neural circuit dynamics of behavior. A small number of investigators have reported the use of Neuropixels 1.0 in humans, revealing insights into neuropathologies[15], speech production[3], language[4,16], and cognition[16]. However, this probe's 1cm length prevents multi-region recordings in larger animals and doesn't allow access to deep structures without resecting the overlying cortex[16]. To address these limitations, a longer version—Neuropixels 1.0 NHP—was developed for use in non-human primates[17–19] (Fig. 1A). The longer of the two variants of this probe features 4,416 recording sites along a 45 mm shank, with 384 selectable channels, enabling dense, multi-site recordings in large brains (Fig. 1B). With cross-sectional dimensions of 125 µm in width and 122 µm in thickness, the Neuropixels 1.0 NHP probe is significantly smaller than conventional human intracranial electrodes such as stereoelectroencephalography (sEEG) electrodes or Utah multielectrode arrays (Fig. 1C), permitting substantially higher spatial resolution while minimizing tissue displacement.

Handling the longer probe is technically challenging and necessitates a workflow that ensures greater surgical accuracy with cortical penetration than the original 1 cm long probe. Here, we report our methods and results for the intraoperative use of the 1.0-NHP probes in nine human participants undergoing intracranial procedures for clinical purposes. We report technical innovations for probe sterilization, custom-designed instrumentation, and intraoperative handling, and describe lessons learned to optimize signal quality and data yield in the human operating room.

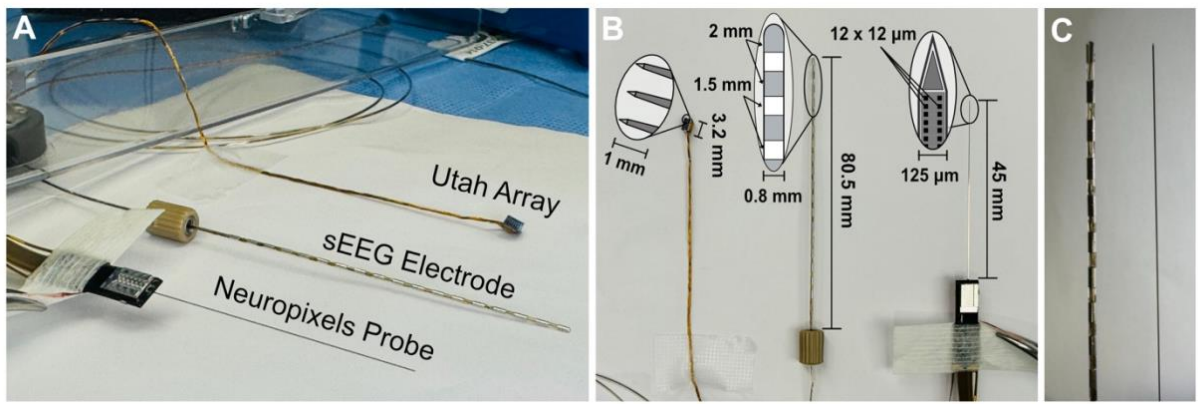

**FIG. 1. Comparison of the Neuropixels 1.0–NHP Long probe with other intracranial electrodes.**
**A:** The Utah array, sEEG lead, and Neuropixels 1.0–NHP Long probe displayed side by side.
**B:** Top-down view of the electrodes shown in panel A, with magnified diagrams of each device's electrode channel configuration. The Utah array contains an 8 × 8 channel grid within a 3.2 × 3.2–mm form factor. The

DIXI sEEG electrode has a diameter of 0.8 mm and 5–18 cylindrical contact sites. The Neuropixels 1.0–NHP Long probe measures 125 μm in width and 122 μm in thickness, with electrode contacts measuring 12 × 12 μm, arranged at a pitch (separation between contacts) of 103 μm (columns) and 20 μm (rows).
**C:** Side-by-side comparison of the sEEG lead and the Neuropixels probe, highlighting the markedly smaller footprint of the Neuropixels device.

## Methods
### Patient Selection and Ethics Approval
The study was reviewed and approved by the UC Davis Investigational Review Board (protocol #2057788). Patients were approached to participate in research by a senior investigator (DMB), who discussed the risks and benefits of participation. Consent was obtained at a separate clinical encounter by a clinical research coordinator. Participants had their questions answered and confirmed their participation again the day of the procedure. Participants are repeatedly informed that their participation, or lack thereof, will not impact their clinical care.

### Electrophysiological and Behavioral Recording Equipment
Neuropixels 1.0 NHP long probes (IMEC, Leuven, Belgium) (Fig. 2A) were modified to include touchproof connector leads for ground and reference (SleepSense Safety DIN to Key Connector Adapter, PS036), which were cut and soldered to the probe's ribbon cable[1] (Fig. 2B). Following electrical validation using built-in tests from OpenEphys version 1.01 (https://open-ephys.org/gui), the probes were packaged into modified sterile trays (base, lid, mat 6 × 2.5 × 0.75; Duraline Biosystems, item no. A-CP614) (Fig. 1C). Tray modifications included two specific cutouts: (1) a large square recess to accommodate the probe shank, and (2) a smaller rectangular space to house the ribbon cable connector and the soldered junctions for ground and reference leads (Fig. 1C).

Recordings were performed using a custom-built mobile computational "rig" consisting of two computers (Fig. 3A and C). The first computer was dedicated to high-throughput acquisition of neural signals from the Neuropixels probe running both OpenEphys version 1.01 and SpikeGLX software (https://billkarsh.github.io/SpikeGLX/), while the second managed behavioral task presentation and generated synchronization signals to align electrophysiological data with external analog inputs (Fig. 3C). For awake procedures, a compact task cart incorporating a speaker and monitor was positioned near the patient (Fig. 3A and B). Audio was recorded using a cardioid vocal microphone (Shure SM58-CN, Niles, USA) (Fig. 3B) and a Tascam DR-40X audio recorder (TASCAM, Tokyo, Japan) (Fig. 3C). The data acquisition system incorporated a National Instruments PXIe-1083 5-slot chassis and a BNC-2110 shielded connector block (National Instruments, Austin, USA) to interface the Neuropixels headstage and external peripherals (Fig. 3C).

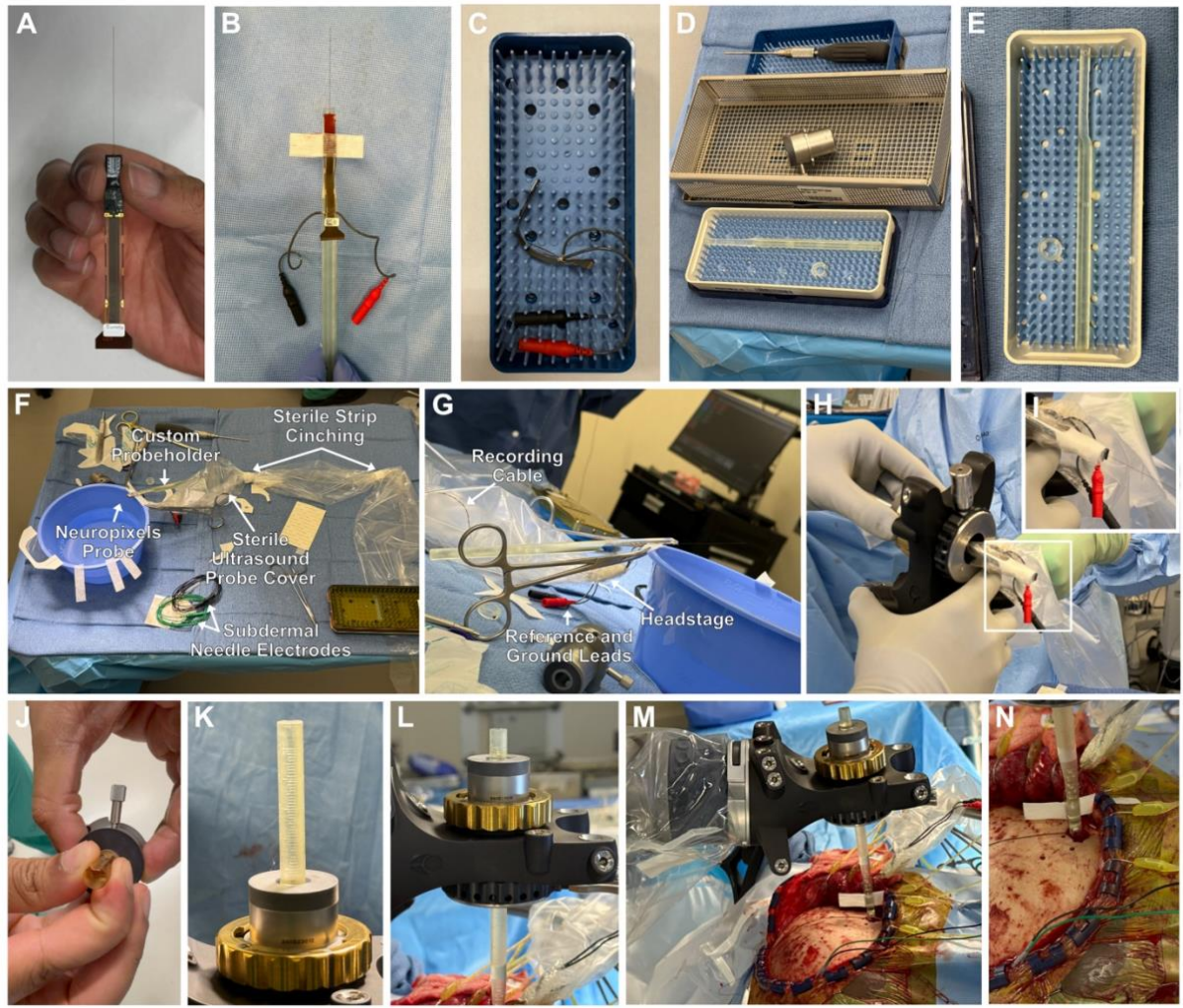

**FIG. 2. Workflow for intraoperative Neuropixels recordings during open craniotomy procedures.**
**A:** Example of a Neuropixels 1.0–NHP Long probe showing the ribbon cable connector.
**B:** Functional electrode assembly showing ground and reference leads soldered to the ribbon cable. Sterile strips are placed at locations optimized for secure gripping with mosquito clamps.
**C:** Neuropixels probe positioned within a sterilization tray. The ribbon cable and ground wires are arranged so that the probe base is wedged between the tray ribbing to minimize movement during sterilization.
**D:** Supporting equipment for intraoperative Neuropixels recordings. *Top:* Sterilized Torx T5 screwdriver. *Middle:* Custom 3D-printed anti-rotation guide tube compatible with the Globus surgical robot. *Bottom:* Custom 3D-printed Neuropixels probe holder with stop cap.
**E:** Close-up of the sterilized custom 3D-printed probe holder and stop cap.
**F:** Fully sterilized recording assembly, with the probe mounted in the holder and the headstage and recording cables enclosed in sterile plastic.
**G:** Enlarged view of the probe secured in the holder and clamped to the sterile bowl containing sterile saline.
**H:** Sterile assembly inserted into the anti-rotation guide block and mounted on the detachable arm of the Globus surgical robot.
**I:** Enlarged view of the probe within the guide tube during insertion via the surgical robot.
**J:** Close-up of the custom probeholder seated within the custom Globus reducing tube, showing the complementary anti-rotation channel.
**K:** Depth-etched markings on the probe indicating the cortical entry point; markings are spaced at 1-mm intervals.

**L:** Depth markings indicating the final insertion depth achieved during recording.
**M:** Full view of the probe in situ, showing the surgical robot, insertion trajectory, and placement of the ground and reference leads, which were inserted into the scalp and irrigated periodically to prevent tissue desiccation.
**N:** Enlarged view of the probe seated within the craniotomy burr hole.

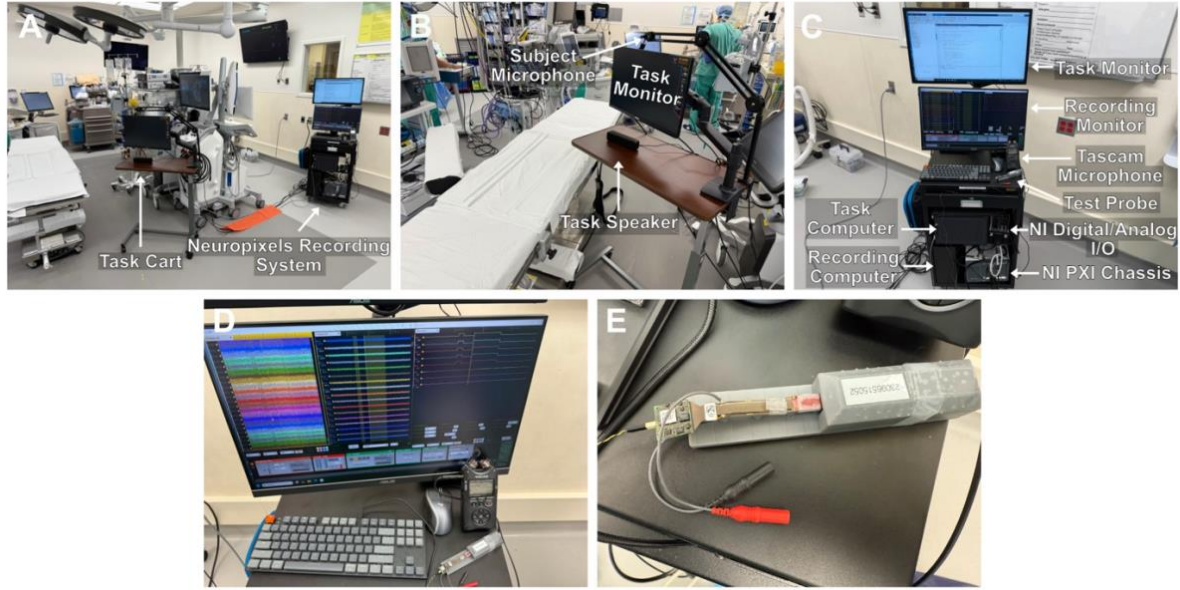

**FIG. 3. Neuropixels operating room (OR) recording system.**
**A:** OR setup showing the Neuropixels recording system and task cart positioned adjacent to the operating table.
**B:** Enlarged view of the task cart in position next to the operative field.
**C:** Close-up of the recording system computers and monitors.
**D:** Example of the recording software interface during testing, with a validation probe placed inside its protective case. Data synchronization pulses are visible in the upper right corner of the monitor.
**E:** Test probe in its custom protective housing, used for system validation and debugging prior to intraoperative recordings.

## Custom Surgical Equipment

The commercially available stereotactic holders (also referred to as "rods") provided by IMEC were not originally designed for integration with standard neurosurgical workflows. These holders are machined from aluminum, which is not accepted by all modern hospital sterilization departments. This necessitates ethylene oxide (EO) sterilization (which is logistically complex and costly), thereby preventing easy reuse and hindering widespread clinical adoption. Additionally, the stock holders fail to center the Neuropixels probe along its central axis, permitting uncontrolled rotation within a stereotactic apparatus (e.g., the arc of a CRW frame), and do not allow for precise depth advancement without a dedicated micromanipulator.

To overcome these limitations, we designed and fabricated a custom 3D-printed stereotactic holder (Fig. 3D and E) using a Formlabs Form 3B printer with Surgical Guide resin (Class 1 Medical Device in compliance with EN ISO 13485, 14971, 10993-5, and 10993-10). This holder offers several key advantages over the stock stereotactic holders[20]: (1) it is constructed from a sterilizable and reusable material, (2) it centers the Neuropixels probe's shank along the device's axis for consistent targeting, (3) it incorporates an anti-rotation channel to prevent twisting about the Z-axis (i.e., the axis pointing down into the cranium) during insertion (Fig. 2J), and (4) it features millimeter-scale depth markings to facilitate controlled and verifiable advancement of the probe into brain tissue (Fig. 2K and L).

Additional custom components were developed to support different surgical workflows. For craniotomies incorporating the ExcelsiusGPS® robotic navigation system[21–23] (Globus Medical, Audubon, USA), we developed a custom guide tube with a matching anti-rotation channel (Fig. 2D and J). For deep brain stimulation (DBS) procedures using the CRW stereotactic frame, a custom DBS guide block was 3D printed (same materials as the holder) to accommodate the probe and its electrical connectors (Fig. 4A, 4L, and 4N–T). This guide block is also compatible with the ROSA One® Brain (Zimmer Biomet, Warsaw, USA). Because Neuropixels' connecting cables are not re-sterilizable, a 3D-printed sterile "straw" insert—compatible with the custom DBS guide block—was developed with a channel to safely pass the electrical cable into the sterile field (Fig. 4A, 4C, and 4L). To minimize the risk of mechanical failure during both surgical modalities, a stable 3D-printed support frame was also designed, allowing the assembled sterile Neuropixels system to rest securely prior to probe insertion (Fig. 4A and L). All tools were printed using the same Surgical Guide resin as the custom stereotactic holder.

Design files for the stereotactic holder, guide tube, guide block, and straw insert have been provided exclusively for noncommercial academic research use [https://zenodo.org/records/17633875][24].

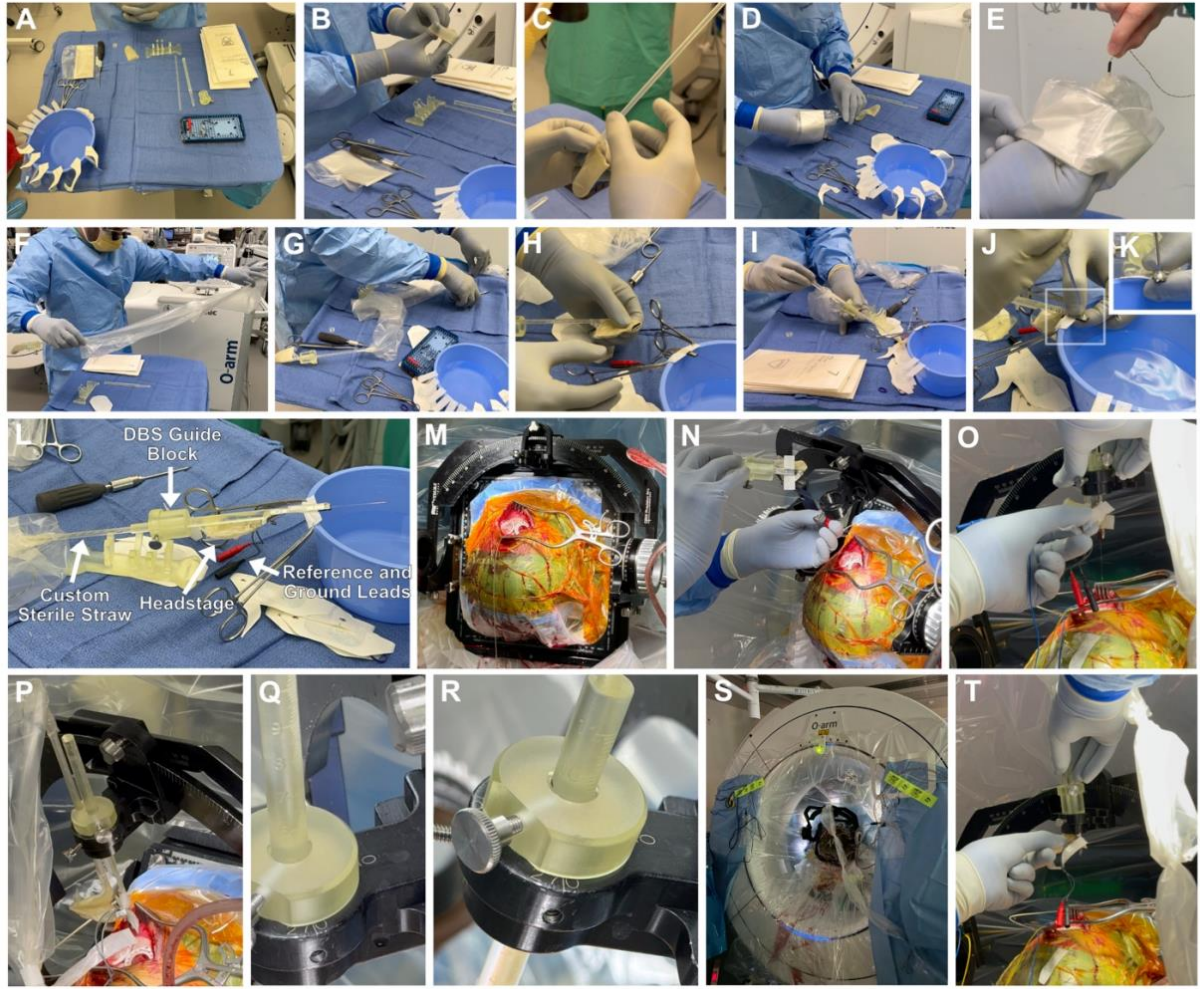

**FIG. 4. Workflow for sterile Neuropixels assembly and intraoperative recordings during deep brain stimulation (DBS) implantation.**

 **A:** Required sterile components arranged on a side table, including the sterilizable Torx T5 screwdriver, custom 3D-printed DBS guide block, custom probeholder, horizontally and vertically cut Tegaderm sheets, size 8½ transparent surgical glove, sterile snaps, scissors, sterile strips, probe cover bag, 3D-printed cable conduit (4.7-mm inner diameter "straw"), and two subdermal needle electrodes.
 **B:** Preparation of the headstage cover: the thumb is cut from the sterile surgical glove, inverted by a sterile team member, and used to enclose the nonsterile headstage to maintain sterility.
 **C:** Introduction of the sterile straw into the glove thumb portion, which is then sealed with Tegaderm to secure the junction.
 **D:** The straw is passed through the custom DBS guide block, with the probe cover bag affixed to the smaller-diameter side of the block. A small opening is made in the bag to permit passage of the straw, which is then sealed in place with Tegaderm.
 **E:** The nonsterile researcher advances the ribbon cable through the straw while the sterile researcher maintains sterility using the inverted probe cover bag.
 **F:** The cable head is grasped within the glove thumb and enclosed within the probe cover bag before being connected to the headstage inside the thumb pouch.
 **G:** The probe cover bag is reinforced with sterile strips at two locations to reduce strain on the headstage–cable interface.
 **H:** After securing the Neuropixels probe following the same procedure used in craniotomy surgeries, a small opening is made in the glove to permit passage of the ribbon cable, which is then connected to the headstage and sealed within the glove using Tegaderm.

**I:** The custom probeholder is inserted into the DBS guide block and aligned with the Neuropixels probe.

**J:** The Neuropixels probe is secured to the probeholder and the sterile assembly using the sterilized Torx T5 screwdriver.

**K:** Close-up view of the probeholder and set screw, showing correct alignment and engagement.

**L:** Completed sterile Neuropixels–DBS assembly ready for intraoperative use.

**M:** Intraoperative view showing bilateral burr holes. The contralateral burr hole (nonrecording target, left hemisphere) contains a standard DBS electrode, while the recording target burr hole (right hemisphere) is prepared for introduction of the sterile Neuropixels assembly.

**N:** Introduction of the sterile Neuropixels assembly into the operative field, with ground and reference leads passed through the CRW frame first.

**O:** The Neuropixels probe and DBS block positioned within the CRW frame, with the ground and reference leads held taut and secured between the operator's fingers.

**P:** Advancement of the Neuropixels probe to the cortical surface.

**Q:** Depth markings visible at the cortical entry site prior to insertion, with the target depth determined from preoperative imaging and verified intraoperatively by multiple team members.

**R:** Final depth position of the Neuropixels probe after insertion.

**S:** Operative field during ongoing electrophysiological recordings, with optional contralateral stimulation.

**T:** Withdrawal of the Neuropixels probe and removal of the DBS block from the CRW frame following completion of recordings.

## Preoperative Preparation

Two to three days prior to each surgery, the recording rig was electrically tested using a test probe housed in a custom 3D-printed protective case (Fig. 3E). Surgical trajectories were reviewed in advance, and appropriate experimental paradigms were selected based on the planned operative approach and anticipated recording conditions.

Experimental paradigms fell into two categories depending on whether the surgery would be conducted under general anesthesia or while the patient was awake. For asleep cases, patients were presented with auditory stimuli consisting of speech in a familiar language (English), an unfamiliar language (Russian), and periods of silence. For multilingual patients (e.g., English–Spanish bilinguals), speech in all three languages was presented. In awake cases, patients were asked to perform simple tasks such as repeating short sentences or executing basic movement commands to allow for task-aligned neural recordings.

## Sterile Surgical Procedure

All Neuropixels recordings were performed intraoperatively during clinically-indicated neurosurgical procedures. In all cases, custom sterile protocols were used to maintain probe sterility and ensure mechanical stability during deployment. In craniotomy recordings, probe trajectories were chosen in areas destined for resection (e.g., temporal neocortex / hippocampus in an anterior temporal lobectomy). In DBS recordings, trajectories were placed across the precise trajectory of the DBS cannula.

Given the length of the probe, precise control of its trajectory and final position within the cortex was essential. Our custom stereotactic holder—used in conjunction with the complementary custom anti-rotation bushing for craniotomy cases performed with the Globus surgical robot, or with a custom DBS guide block for DBS procedures performed using the CRW or ROSA robotic systems—minimized angular deviation of the probe during both insertion and withdrawal.

No more than 30 minutes of time was dedicated to research. This includes both introducing research equipment to the sterile field and intracranial recordings.

**Craniotomy Surgeries**

For participants NPX01–NPX05, the neuropixels probe was inserted after completion of the clinically indicated craniotomy and durotomy. For participants NPX06–NPX08, we placed the Neuropixels probe through a burr-hole craniostomy and a small durotomy (Fig. 2M). This craniostomy was then incorporated into the final craniotomy used for resection.

A sterile assembly first described in Coughlin, Muñoz, Kfir et al. 2023 was adapted for use with the Neuropixels 1.0-NHP (Fig. 2F). Briefly, after removing the Neuropixels probe from the sterilization box, we applied Steristrip dressings (3M, St. Paul, USA) on the probe, and affixed it to a bowl filled with sterile saline with mosquito snaps (Fig. 2G). This had the advantage of (1) stabilizing the probe at a comfortable working height, and (2) minimizing the chance of damaging the probe during the sterile assembly process (since it could be clearly seen against a liquid background). Next, an ultrasound probe cover was introduced and then the Neuropixels headstage (already connected to the ribbon wire) was placed within the inside of a bag. A small cut was made into the ultrasound bag, and the probe was then connected to the headstage through the sterile plastic. The interface was then sealed with multiple Tegaderm dressings (3M).

The probe was mounted onto a custom 3D-printed stereotactic holder and secured by tightening a super–corrosion-resistant cup-point set screw (316 stainless steel, M2.5 × 0.45 mm thread, 3 mm length; McMaster-Carr, Elmhurst, USA) with a sterilizable Torx T5 screwdriver (Part #86.1605, Millennium Surgical Corp., Narberth, USA), ensuring stability and preventing probe movement[24]. The custom stereotactic holder was then mounted within a custom bushing designed to mount within the end-effector of the Globus robot (Fig. 2H and I). This bushing was designed to accommodate the anti-rotation design of the holder (Fig. 2J).

Upon beginning research recordings during the operation, the end-effector was attached to the robot arm and the robot introduced into the sterile field. Once the arm was in its final location, the end-effector was removed, and then the bushing/Neuropixels complex was assembled, before being re-attached to the robot (Fig. 2M). We note that the design of the robot prevents the surgeon-driven or collinear advancement of the end-effector while instruments are placed through the bushing.

Immediately prior to probe insertion and initiation of electrophysiological recording during the 30-minute research protocol, a standardized noise-reduction procedure was performed. All operating room equipment that could be safely powered down was either turned off or switched to battery operation. This included the Globus robotic platform, monopolar and bipolar electrocautery devices, and the anesthesia infusion pump. All mobile phones in the operating room were turned off or placed in airplane mode with Bluetooth disabled for the duration of the recording.

**Deep Brain Stimulation Surgeries**

Our standard surgical workflow for DBS uses the CRW frame and ROSA robot. Since the ROSA provides the same holder, here we describe the workflow for the CRW frame. The sterile assembly involved several components. We developed a custom 3D-printed Neuropixels DBS guide block designed to replace the standard guide block (component CRWPGB) that sits in the Precision Arc holder. We modified the standard guide block with two innovations: a hole designed to accommodate a 3D printed flexible tunneler sheath (4.7 mm inner diameter), and a modification designed to accommodate the anti-rotation device of the holder (Fig. 4L and P).

Since it could be unsafe to remove the guide block holder from the CRW during the operation (i.e., remove the arc to place the probe, and then re-attach the arc) we modified the workflow developed for the Globus approach. Rather than placing the headstage directly into the ultrasound probe cover, it was placed into the "thumb" cut from a standard size 8.5 latex glove (Fig. 4B). A custom 3D printed tunneler sheath ("straw") was then introduced (Fig. 4C). One end of the sheath was taped to the glove containing the headstage, and the other end was subsequently passed through the DBS guide block. The ultrasound probe cover was then secured to the open end of the sheath (Fig. 4D), allowing the electrical cable to be passed intraoperatively while maintaining the sterile field (Fig. 4E). The cable head was grasped within the glove thumb, while the remaining length of the wire was enclosed within the probe cover bag before being connected to the headstage inside the thumb pouch (Fig. 4F). The probe cover bag was then reinforced with sterile strips to minimize strain on the headstage–cable interface and maintain cable stability during manipulation (Fig. 4G). Thereafter, a small hole was cut in the glove and the headstage was affixed to the probe (Fig. 4H).

The rest of the setup was as described above for the Globus workflow: the ribbon cable was routed through the glove and connected to the headstage maintaining the sterile field. Multiple strips of Tegaderm were used to reinforce the headstage-cable-sterile bag interface and sterile strips were used to further stabilize the probe to the 3D printed stereotactic holder (Fig. 4I–L).

After clinically indicated contralateral burr holes were made and the DBS electrode was placed into the nonrecording target (FIg. 4M), the sterile Neuropixels assembly was introduced into the operative field (Fig. 4N and O). Reference and ground leads were first passed through the CRW Precision Arc holder (Fig. 4N) and carefully tensioned as the Neuropixels probe with guide block was slid into position (Fig. 4O). The reference and ground leads were then secured away from the probe using Tegaderm or sterile strips. Research recordings adhered to the same strategies for electrical noise reduction as those described for craniotomy-based procedures, with the exception that contralateral electrical stimulation was selectively applied during recordings to assess the feasibility of using long Neuropixels to measure potential stimulation-related neural effects.

**Electrophysiological Data Collection and Processing**

The Neuropixels probe was configured with a channel mapping tailored to each patient's anatomy along the trajectory selected for the clinically indicated surgical procedure. Channel maps were determined by balancing two primary considerations: (1) maximizing the span of gray matter sampled along the planned probe trajectory, and (2) optimizing signal density and quality to enable reliable single-unit isolation. External reference and ground leads were connected to subdermal needle electrodes which were then inserted into scalp tissue. An external reference was used for probes NPX01–NPX07, whereas a tip reference configuration was used for NPX08–NPX09. Raw neural signals were preprocessed using SpikeInterface[25], with motion correction of nonstationary neural signals performed using the LFP-based DREDge algorithm[26], and subsequently spike-sorted with Kilosort 4[1,27]. Automatically detected clusters were subsequently manually curated in Phy[28] by an experienced electrophysiologist. Clusters that could not be further resolved into single units were classified as multi-unit activity.

## Results

Nine participants (5 male, 4 female; mean ± SD age 37.6 ± 12.5 years) underwent intraoperative implantation with Neuropixels 1.0–NHP long probes during clinically indicated neurosurgical procedures for medically refractory epilepsy, including resections (n = 6) and DBS (n = 3). Two recordings were performed in awake patients and seven under general anesthesia. Case details are in Table 1; successful neural recordings were obtained in all cases.

No probe breakages occurred during implantation or explantation. Recording durations ranged from 5 to 21 minutes (intentionally limited to minimize extending surgery lengths), with insertion depths spanning approximately 5 mm to 40 mm from the cortical surface. In most cases, the probes were inserted at almost the full length of the probe, at a depth of 40 mm. Multiple recording configurations were used, yielding data across spans ranging from ~3.8 mm to 30 mm. We found that smaller burr holes resulted in much less brain motion relative to the probe (Fig. 5A). Targeted brain regions included: (1) the non-dominant hand area of the postcentral gyrus, (2) the non-dominant posterior middle frontal gyrus, (3) the hippocampus accessed through intact middle temporal gyrus (Fig. 5B), and (4) cingulate cortex. We demonstrated that complex speech behavior tasks could be performed during the awake behaviors (Fig. 5C). All patients recovered without any adverse events or complications related to the use of the Neuropixels probes.

We developed a checklist to review after the probe had been inserted into the brain, in order to optimize recording quality. Reference and grounding electrodes were checked throughout the procedure to ensure stable placement (i.e., not pulling out of scalp tissue). We found it helpful to periodically moisten the tissue adjacent to the grounding electrodes with sterile saline, though care had to be taken to avoid electrical current shorts caused by pooling saline. We also discovered additional electromagnetic noise sources in our operating room from various clinical equipment, including the O-arm, Globus robot, monopolar cautery, and bipolar cautery. We therefore unplugged these devices during recordings. We found that unplugging the intravenous pump used for anesthesia so that it ran on batteries also decreased electrical noise. Cell phones also generated interference, and we placed them in "airplane mode" with Bluetooth turned off during recordings.

| ID# | Sex | Age | Procedure | Awake/Asleep | Hemisphere | Recording Target(s) | Stim? (Y/N) | Anesthesia | Surgical System Used | Probe Holder Used | Insertion Depth | Recording Span | Channel Map | Task/Stimuli | Recording Duration |
|---|---|---|---|---|---|---|---|---|---|---|---|---|---|---|---|
| NPX01 | F | 21 | Epilepsy Tissue Resection | Awake | Non-dominant | Hand Area of Post-Centreal Gyrus | No | N/A | Globus | Original Aluminum | ~5 mm | ~3.84 mm | Dense 2 rows | Movement + speech | ~9 min |
| NPX02 | M | 25 | Epilepsy Tissue Resection | Asleep | Non-dominant | Hippocampus | No | 1.0 MAC of Sevoflorane and remi-fentanyl | Globus | Original Aluminum | ~40 mm | ~7.68 mm | Tetrodes | Listening (Eng) | ~8min |
| NPX03 | F | 23 | Epilepsy Tissue Resection | Asleep | Non-dominant | Hippocampus | No | 0.5MAC of Sevoflorane and remi-fentanyl | Globus | Original Aluminum | ~41 mm | ~7.68 mm | Tetrodes | Listening (Eng) | ~5 min |
| NP04 | F | 39 | Epilepsy Tissue Resection | Asleep | Non-dominant | Hippocampus | No | sevoflorane and fentanyl | Globus | Original Aluminum | ~40 mm | ~7.68 mm | Dense 1 row | Listening (Eng, Rus) | ~12 min |
| NPX05 | M | 50 | Epilepsy Tissue Resection | Awake | Non-dominant | Posterior Part of the Middle Frontal Gyrus | No | N/A | Globus | Original Aluminum | ~35 mm | ~30 mm | Sparce 1 row | Critical information | ~5.5 min |
| NPX06 | M | 63 | Epilepsy Tissue Resection | Asleep | Non-dominant | Hippocampus | No | 5 micrograms Sevoflurane 75 micrograms per kilo of propofol | Globus | Custom 3D Printed | 38 mm | ~7.68 mm | Dense Linear Map | Listening (Eng, Span, Rus) | 21.6 mins |
| NPX07 | F | 36 | DBS | Asleep | Non-dominant | Middle Frontal Gyrus | Yes | Remifentanyl, sevoflorane | CRW | Custom 3D Printed | ~8 mm ~18 mm | ~7.68 mm | Dense 1 row | Stimulation | ~18.8 mins |
| NPX08 | M | 39 | DBS | Asleep | Non-dominant | Cingulate | Yes | Remifentanyl, sevoflorane | CRW | Custom 3D Printed | 40 mm | ~3.84 mm | Dense 2 rows | Stimulation | 17 min |
| NP0X9 | M | 44 | DBS | Asleep | Non-dominant | Superior Frontal Sulcus | Yes | Remifentanyl, sevoflorane | ROSA | Custom 3D Printed | 19 mm | ~3.84 mm | Dense 2 rows | Stimulation | 20 min |

**TABLE 1. Patient Characteristics, Surgical Details, and Intraoperative Neuropixels Recording Conditions**

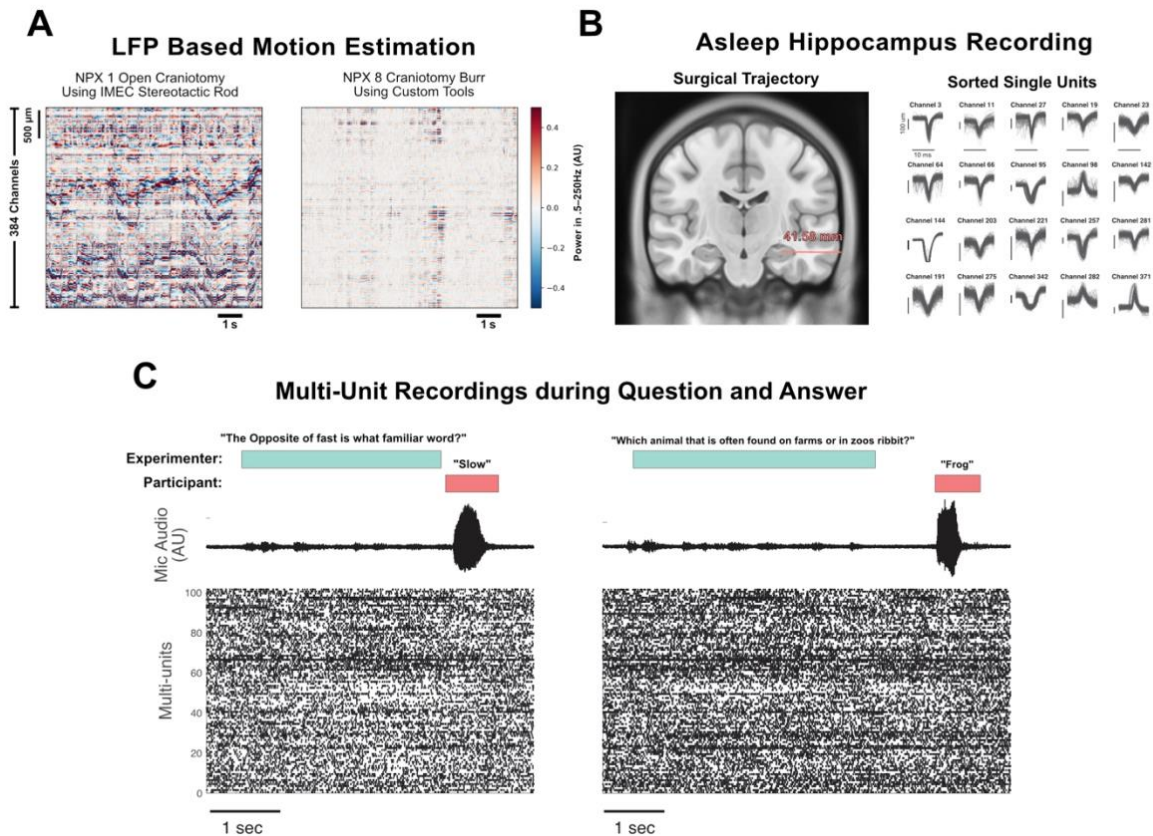

**FIG. 5. Representative intraoperative neuropixels recordings across behavioral and pathological states.**
**A:** Representative example of brain tissue motion relative to the probe during an awake open craniotomy recording (NPX01), illustrated by local field potential (LFP) activity displayed as z-scored power in the 0.5–250 Hz band (left). For comparison, the right panel shows LFP activity recorded through a burr-hole craniostomy (NPX08), demonstrating substantially reduced tissue motion relative to the probe.
**B:** Left: coronal view of the planned surgical and recording trajectory targeting the hippocampus during a temporal lobectomy (NPX05). Right: representative well-isolated single units obtained from this recording.
**C:** Multiunit activity recorded during an awake speech task involving a question–answer paradigm (NPX01).

# Discussion

We developed methods for safely recording from patients using the recently developed large-brain-optimized Neuropixels 1.0 NHP long probes. These methods allowed recordings in patients undergoing resections and those undergoing DBS procedures. This capability enables large-scale, high-temporal resolution recordings from both healthy and pathological tissue, at depths never before described in the human literature. In addition, the ability to stimulate contralateral subcortical targets while simultaneously recording from deep structures opens new avenues for mechanistic and translational neuroscience research.

### Practical Considerations: Sterility, Workflow, and Signal Reliability

During the development and refinement of these methods, we identified several factors that influence signal quality and best practices to prevent failure modes during recordings. These potential pitfalls fall broadly into three categories: (1) logistical—ensuring that equipment is sterilized, recording instruments are assembled correctly, and procedural steps are clearly documented; (2) technical—preventing probe breakage during insertion and removal, protecting associated equipment, and minimizing brain motion relative to the probe; and (3) electrical—minimizing signal interference and ensuring that all connections are secure.

### Maintaining Sterility

Maintaining sterility required particular care. The Neuropixels probe itself can be sterilized; however, the headstage and flexible cable—which must be connected for recordings—can only be purchased pre-sterilized with no protocol for resterilization. We addressed this issue by placing the nonsterile headstage, ribbon cable, and flexible cable in sterile plastic and steri-strips prior to probe insertion. We found it easiest to assemble the full recording system on a sterile side table prior to starting the operation (e.g., during induction). We positioned the probe securely in the stereotactic holder (Figs. 2F–I and 4J–L) before mounting it to either the robotic arm (Fig. 2H and I) or DBS block (Fig. 4L) and introducing it into the surgical field.

### Mitigating Brain Motion to Improve Recording Stability

Physiological and behavioral factors—including respiration, cardiac pulsation, coughing, and speech—induce appreciable brain motion in surface recordings. This movement caused neurons to travel dramatically in relation to the tight spatial resolution of the Neuropixels (Fig. 5A left). To mitigate this effect, we altered our technique by placing the probe through a burr-hole that would eventually be incorporated into the final craniotomy (Fig. 2M and N). . Our data suggests that reducing the amount of exposed brain substantially reduces neuronal displacement during recordings, thereby improving signal stability (Fig. 5A right).

### Custom 3D-Printed Surgical Tools: Design and Sterilization Challenges

To support these methods, we designed and 3D-printed several custom surgical tools. These provided advantages over vendor-provided options by centering the probe shank along the surgical trajectory, preventing unwanted rotation during insertion, and allowing sterilization and reuse. When designing and manufacturing these tools, attention to mechanical tolerances and packaging for sterilization was essential. Although the 3D printing target tolerance was approximately ±0.002 mm, batch-to-batch variation in resin properties can lead to dimensional shifts of up to 0.01 mm. Consequently, test prints with incremental adjustments (on the order of 0.001 mm) are recommended for critical dimensions such as the dovetail and inner/outer diameters. In addition, resin-based tools must be adequately supported during sterilization to prevent deformation caused by heavier instruments stacking on top of them during cooling. Modifying sterile trays to physically isolate each component and prevent stacking during sterilization is an effective solution (Fig. 2E). These seemingly minor details are critical, as even small deviations in tolerancing or alignment can potentially result in substantial errors in probe trajectory.

### Anesthetic Considerations for Neural Signal Quality

Another consideration was the choice of anesthetic agents. Our clinical workflow is to provide 0.5MAC of sevoflorane and remi-fentanyl as maintenance anesthetic, in operations with general anesthesia in which recordings are needed (e.g., DBS for Parkinson's disease). We provide further recommendations to our anesthesiology colleagues to avoid use of long-acting benzodiazepenes during the entire operation, and that propofol not be used after induction. We suggest that future work should include systematic study of anesthesia pharmacokinetics and timing relative to Neuropixels recordings, ideally in collaboration with clinical anesthestiologists and pharmacology experts.

**Probe Integrity and Safety**

A major concern with the use of Neuropixels probes in humans is the possibility of in situ breakage. Prior work with shorter probe designs has documented this risk[15,20,29]. In our experience, we have not yet encountered breakage when using the longer version of the probe. It is possible that the additional length increases the distance between the insertion site and the probe base, thereby attenuating lateral and torsional forces at the region most prone to failure. Nonetheless, the risk cannot be excluded, and continued monitoring across larger cohorts will be essential to determine the true safety profile of these devices.

All intracranial insertion electrodes, whether used acutely or chronically, inevitably produce some degree of microdamage to brain tissue[30]. Although shorter Neuropixels probes—sharing the same cross-sectional dimensions as the longer variants—have been shown to cause several orders of magnitude less observable tissue disruption than conventional sEEG electrodes[29], further studies are necessary to evaluate the long-term effects of inserting these longer probes into brain tissue.

**Advantages and Implications for Neuroscience and Clinical Translation**

The principal advantage of our protocol is the ability to safely and simultaneously record from many neurons simultaneously across cortical and subcortical structures—without resecting overlying tissue—with millisecond temporal resolution and submillimeter precision. For instance, we were able to record from the hippocampus without first resecting the temporal neocortex[16], or the cingulate sulcus through a standard DBS-lead trajectory. The custom equipment we provide reduces barriers to adoption at other sites and enables investigation of both healthy and diseased neural tissue while preserving its connectivity to the rest of the brain. We anticipate that these methods will be a valuable contribution to both basic and translational neuroscience and may ultimately inform more precise clinical interventions.

# Conclusions

The Neuropixels 1.0–NHP Long probes were successfully used to obtain high-quality intraoperative recordings. The demonstrated new capabilities provided by these longer probes relative to previous human intraoperative Neuropixels recordings are the ability to record in deeper brain structures and across multiple brain areas. In contrast to prior reports using shorter probes, no probe breakages were observed. To facilitate broader adoption, we developed custom tools and methods that reduce technical barriers for both clinicians and researchers. As these approaches are adopted more widely and the technology continues to advance, we anticipate they will enable new insights into human neuroscience and neuropathology and may ultimately contribute to the development of more precise clinical interventions.

## Acknowledgments

The authors thank Jessica Beatty and Melanie Ann Peters for their outstanding support as clinical research coordinators. Their contributions were essential to the successful conduct of these studies. This publication is based on research supported by The Pershing Square Foundation, Bill Ackman, and Neri Oxman (MIND Prize to S.D.S.). This research was also supported by: Schmidt Science Fellows in partnership with the Rhodes Trust, the ASEE eFellows Program, a Burroughs Wellcome Fund Postdoctoral Diversity Enrichment Program Award (D.E.B); an NIH R01 NS134410-01A1 (A.C.P); the Burroughs Wellcome Fund CASI, and the Searle Scholars Program (S.D.S.). The content of this work is solely the responsibility of the authors and does not necessarily represent the official views of Schmidt Science Fellows, the Rhodes Trust, or the United States Government.